\documentclass{aa}  
\usepackage{placeins}

\usepackage{graphicx}
\usepackage{mathabx}

\usepackage{txfonts}
\usepackage{xcolor}

\newcommand{\rebound}{\textsc{REBOUND}\xspace}
\newcommand{\betapic}{\textsc{$\beta$} Pic\xspace}

\begin{document}

   \title{Exocomets of $\beta$ Pictoris}
   \subtitle{II: Two dynamical families of exocomets as simulated with \rebound.}

   \author{K.P. Jaworska
          \inst{1}
          \and H.J. Hoeijmakers
          \inst{1}
          }

        \institute{
    Lund Observatory, Division of Astrophysics, Department of Physics, Lund University, Box 118, 221 00 Lund, Sweden
             }

   \date{}

  \abstract
  {}
   {We investigate the dynamical evolution of particles in the \betapic system to determine likely formation pathways to the present-day observed exocomet populations. We aim to relate these results to similar studies recently carried out since the discovery of the inner planet \betapic c.}
   {We simulate the \betapic system using the non-symplectic adaptive N-body integrator IAS15 in \rebound. We seed the system with over 100`000 mass-less test particles that evolve for 25 Myr, and adopt initial conditions and a particle distribution that closely matches similar simulations in recent literature. Using IAS15, \rebound resolves close-encounters between test particles and the two gas giants in the system, which is crucial for understanding aspects of the dynamical evolution.}
   {Planet-disk interactions rapidly clear most of the system within 35 AU apart from a region within the orbit of \betapic c, and a region between 20 -- 25 AU. After 10 Myr, exocomets can be sourced continuously from these regions, as well as from the inner edge of the region beyond $\sim$ 35 AU where particles are stable on longer timescales. From the region interior to \betapic c, the exocomets are formed by excitation via mean-motion resonance with \betapic c, obtaining a narrow distribution of radial velocities, consistent with spectroscopic observations. Particles initialized in the outer system may enter onto stargrazing orbits due to disruption by the two gas giants, causing a wider radial velocity distribution, and we propose that this population corresponds to a second dynamical family previously observed via spectroscopy. These particles typically undergo chaotic dynamical evolution for $10^2$ to $10^3$ years after passing the water sublimation limit at $\sim$ 8 AU until reaching the sublimation distance of calcium near 0.4 AU, implying that the two families of exocomets may have different volatile contents. This dynamical simulation depends on the assumed initial conditions, including the system parameters and the morphology of the initial disk, precluding accurate predictions of the absolute rates at which particles are generated from within and from the exterior of the orbits of the two gas giants. However, we predict that it is a general property of this system that a fraction of particles originating from outside the orbit of \betapic b can be scattered onto stargrazing orbits.}
   {}


   \maketitle

\section{Introduction}
Beta Pictoris ($\beta$ Pic) is a young nearby stellar system located at a distance of $19.28 \pm 0.19$ pc \citep{Crifo_1997}, and is estimated to be 23 million years old \citep{mamajek_and_bell_2014}. The system consists of an A-type star surrounded by a large circumstellar debris disk that is seen edge-on from Earth \citep{Smith1984} that has been a subject of interest since its discovery in the early 1980's. It was discovered using the Infrared Astronomy Satellite (IRAS) through its infrared excess \citep{Aumann_1985} and later directly imaged \citep{Smith1984}. The disk was observed to be warped \citep{Burrows1995}, and early on this was recognized as evidence for the gravitational influence of one or multiple unseen exoplanets \citep[e.g.][]{Mouillet1997}. Over a decade later, the $\sim 12$ M$_{\text{J}}$ planet $\beta$ Pic b was finally discovered via direct imaging in a 20-year orbit at a distance of 9.91 AU \citep{Lagrange_2010,Lacour_2021}. Long-term radial velocity follow-up later revealed the presence of the $\beta$ Pic c \citep{Lagrange_2019}, orbiting interior to planet b with a semi-major axis of 2.70 AU and an orbital period of 3.4 years, with a similar high mass as the outer planet. This planet was rapidly followed up and confirmed by high-contrast interferometry observations \citep{Nowak2020}. \\

\noindent The relative proximity and youth of the system make it a valuable laboratory for studies of planet formation environments. Notably, the system is being observed intensively with the newly commissioned James Webb Space Telescope (JWST), enabling unprecedented new observations of \betapic b as well as the disk \citep[e.g.][]{Kammerer2024,Chen2024,Worthen2024}.

\subsection{Infalling exocomets}
Non-photospheric absorption features have been observed in the spectrum of $\beta$ Pictoris since over 35 years \citep{Ferlet1987}. These observations have mainly been made in the cores of the deep stellar Ca II K and H lines at 393.366  and 396.847 nm respectively \citep{edlen1956}, where there is a stable narrow component at a radial velocity close to the systemic velocity attributed to circumstellar gas sourced from the debris disk \citep{Slettebak1983,Hobbs1985}
as well as additional absorbing components that are highly variable, and mostly occur at radial velocities red-shifted by tens of km/s from the narrow component \citep{Ferlet1987}. On account of their redshifts, the bodies sourcing the variable absorption lines are on trajectories towards the star, and these observations were quickly interpreted as absorption from sublimated material released from star grazing planetesimals transiting in front of the star, termed Falling Evaporating Bodies (FEB) or exocomets \citep[e.g.][]{Ferlet1987,Lagrange1988,Beust_1990}. Since the first spectroscopic observations of the \betapic system, many thousands of high-resolution spectra have been obtained, forming a large sample of observations of transiting exocomets from which their statistical properties can be derived \citep[e.g.][]{Kiefer_2014}. Transits of exocomets have also been observed in broad-band light-curves obtained with the Transiting Exoplanet Survey Satellite (TESS) \citep{Zieba2019}.
 
\subsection{Orbital dynamics of exocomets}
The large radial velocities and violent sublimation observed spectroscopically imply a dynamically excited inner system. For this reason, the existence of exocomets was already used as evidence for the presence of gas giants in the inner system, before these were directly detected \citep{Beust_1990}. It is now well known that the orbits of planetesimals in the disk may evolve to eccentric stargrazing orbits due to gravitational interactions with the gas giants \citep[e.g.][]{Beust_2024}.\\

\noindent Since the first discovery of exocomets, studies using dynamical simulations have identified resonant perturbations with an unseen planet as a likely source of exocomet progenitors \citep{Beust1996}. In this scenario, exocomet progenitors are sourced from inner mean-motion resonances (MMR) with the planet (e.g. 3:1 or 4:1) that is itself on an eccentric orbit, gradually increasing their eccentricity until they become stargrazing \citep{Yoshikawa1989}. This model has been invoked to explain a crucial observational property of the infalling exocomet population: The majority of exocomet events are red-shifted, implying a preferred directionality in their infalling trajectories. Resonant orbital evolution with an eccentric planet would naturally result in a directional eccentricity enhancement, determined by the orientation of the orbit of the eccentric planet, and so the mean-motion resonant model has been the leading hypothesis to explain the observed exocomet phenomenon. It was suspected that the orbital characteristics of \betapic b could fit this model \citep{Chauvin2012,Kiefer_2014}, but the discovery of \betapic c as a massive gas giant interior to \betapic b \citep{Lagrange_2019} has made the system more complex than anticipated. A new numerical study was recently carried out by \citet{Beust_2024}, that concludes that the mean-motion resonance model can still explain the exocomet phenomenon, albeit with progenitor particles sourced from mean-motion resonant regions within a pseudo-stable disk interior to the orbit of \betapic c, with an extent of up to 1.5 AU from the star.

\subsection{Dynamical simulations of the \betapic disk with \rebound}
Most previous numerical simulations of the orbital evolution of planetesimals in the \betapic system have been carried out with symplectic integrators \citep[most recently by][]{Beust_2024}. Symplectic integrators offer the advantages of accurate long time-scale stability and high computational efficiency needed to evolve large numbers of particles, but may struggle to resolve close encounters between test particles and the planets. Potentially, the first use of a non-symplectic N-body simulation to model the dynamics of exocomets in \betapic was recently published by \citet{RodetLai2024}, who find that secular evolution of planetesimals in the outer regions of \betapic may result in efficient generation of exocomets after 10 million years - half the age of the system. This simulation was carried out with a much smaller number of particles than \citet{Beust_2024}, and did not make predictions for the distribution of radial velocities of infalling comets, precluding direct comparison with the MMR model. Conversely, the study by \citet{Beust_2024} focuses exclusively on the exocomets that are sourced by the MMR process from the region interior to the orbit of \betapic c, and does not establish that exocomets may be generated from the regions exterior to \betapic b.\\

\noindent In this paper, we do a focused study of the dynamics of exocomets in \betapic with \rebound \citep{Rein_and_Liu_2012}, combining the accuracy of a non-symplectic integrator to resolve close encounters at a scale similar to that achieved by \citet{Beust_2024}, using 133,500 mass-less test particles initialized between 0.5 and 45 AU. Our primary aims are to predict the orbital dynamics of exocomets that exist at the age of the system at around 25 million years of dynamical evolution, the nature/properties of transit events by the population of comets that is observed spectroscopically \citep[e.g.][]{Kiefer_2014}, and to establish a comparison with the dominant MMR theory as recently modified to predict exocomet formation from a debris disk close to the star, interior to 1.5 AU \citep{Beust_2024}. As our principal aim is to carry out a comparison with the results from \citet{Beust_2024}, we typically follow their choices of system parameters and initial conditions in setting up our simulation. These choices include the definition of an initial disk of test-particles that is uniformly distributed in orbital distance, ranging from the inner system starting at 0.5 AU out to an outer distance of 45 AU, with randomly drawn orbital parameters that place the particles on orbits that are close to circular (see Section \ref{sec:methods}). We also adopt one of their solutions for the parameters of the system, originally published by \citet{Lacour_2021} (see Table \ref{tab:initial_conditions}), noting that \citet{Beust_2024} established that their results are largely independent of this choice of solution. We also note that the choices of initial conditions by \citet{Beust_2024} largely match those of \citet{RodetLai2024}, including the choice of solution of the system parameters, meaning that this work enables a direct comparison with both of these previous studies. Finally, we acknowledge that this configuration of the planetesimal disk of \betapic does not correspond to the observed structure of the present-day \betapic disk, in particular the existence of a secondary component in the outer region that is inclined with the orbital plane of the planets by approximately 5$^{\circ}$ \citep{Golimowski_2006}. Our simulation and the simulations that we compare with \citep{Beust_2024,RodetLai2024} therefore do not ascertain whether particles in this inclined disk might also evolve onto  observed statistics \citep{Kiefer_2014}.\\

\noindent This paper proceeds to discuss our methodology in more detail in Section \ref{sec:methods}, and presents the results of the simulation in \ref{sec:results}.

\section{Methods}\label{sec:methods}
N-body simulations of the system were performed using \rebound,  with the non-symplectic integrator \verb|IAS15|. This integrator is well suited for the simulation of exocomets as it can handle close encounters and highly eccentric orbits through implementation of an adaptive step-size. Symplectic integrators are popular because of their speed compared to non-symplectic N-body integrators and their ability to not accumulate energy errors over long timescales \citep{Chambers_1999}. However, \verb|IAS15| has been found to preserve the symplecticity of Hamiltonian systems even better than commonly used symplectic integrators while still being able to simulate non-Hamiltonian systems \citep{Rein_and_Speigel_2015}. The energy error of \verb|IAS15| behaves like a random walk following Brouwer’s law, and this property is deemed optimal by \cite{Rein_and_Speigel_2015}, who successfully use it to simulate cometary close encounters and planet-forming debris disks.

\subsection{Simulation Set-Up}\label{sec:sim_setup}

\noindent With the aim to relate to recent findings of \cite{RodetLai2024} and \cite{Beust_2024} we set up a simulation of the \betapic system in \rebound, together with a disk of mass-less test particles that represent planetesimals. The simulation uses the set of system parameters of the planets and the host star also used by \citet{Beust_2024}, originally measured by \citet{Lacour_2021}, given in Table \ref{tab:initial_conditions}. Additional to the parameters presented in Table \ref{tab:initial_conditions}, the stellar radius and planetary radii we adopt are R$_{\text{star}} = 1.8$ R$_\Sun$ \citep{DiFolco2004} and $\text{R}_{\text{b}} = 1.4 \, \text{R}_{\text{Jup}}$ \citep{Gravity2020, Worthen2024, Ravet2025}. An equal radius was used for planet c such that R$_{\text{c}}=$ R$_{\text{b}}=1.4 \, \text{R}_{\text{Jup}}$.\\

\noindent We populate the system with a disk of test particles, with orbital semi-major axes $a$ ranging from an inner boundary at 0.5 AU to an outer boundary of 45 AU, with a uniform density of 3000 particles per AU, leading to a total of 133,500 mass-less test particles. The particles were initiated with small initial eccentricities $e$ uniformly distributed between 0 and 0.05, and inclinations $i$ uniformly distributed between 0 and 2$^\circ$ respectively. The initial arguments of periastron $\omega$, longitudes of ascending node $\Omega$ and mean longitudes $l$ of the test particles were drawn randomly between 0 and 360 $^\circ$. This follows the setup of the simulation of \citet{Beust_2024} and besides the range in semi-major axes, also follows the choices of \cite{RodetLai2024}. These initial conditions are summarized in Table \ref{tab:initial_conditions2}.\\

\noindent The system was simulated for 25 million years, roughly corresponding to its current age \citep{mamajek_and_bell_2014}. Particles with orbits in the vicinity of \betapic c have relatively short orbital periods and are prone to undergo many close encounters. To resolve close encounters properly, we chose to evolve particles with different maximal time-steps, depending on their position in the system: Particles with a periastron distance greater than 3 AU were simulated using a maximal step of 5 years$^{-1}$, whereas particles with a periastron distance less than 3 AU (either because their orbital parameters evolved during the simulation or because they were initialized on such orbits) were simulated using a maximal time-step of 20 years$^{-1}$. If the periastron distance of a test particle shrunk to within this 3 AU boundary, its simulation was halted, and the particle was reinitialized with its last orbital parameters and continued with the finer timestep for the remaining duration of the simulation, regardless of whether it would move permanently back onto wide orbits beyond \betapic b (the switch from low to high-resolution is non-reversible). The 3 AU criterion was not activated for particles for which the periastron distance shrinks only briefly during a close-encounter with \betapic b, by applying an additional requirement that the position of the particle is less than 5 AU when the difference between the periastron distance and 3 AU limit is evaluated. A test of the sensitivity of the results to the time-step resolution used can be found in Appendix \ref{sec:Appendix_sensitivity_tests}. \\

\begin{figure}[ht!]
    \includegraphics[width=0.47\textwidth]{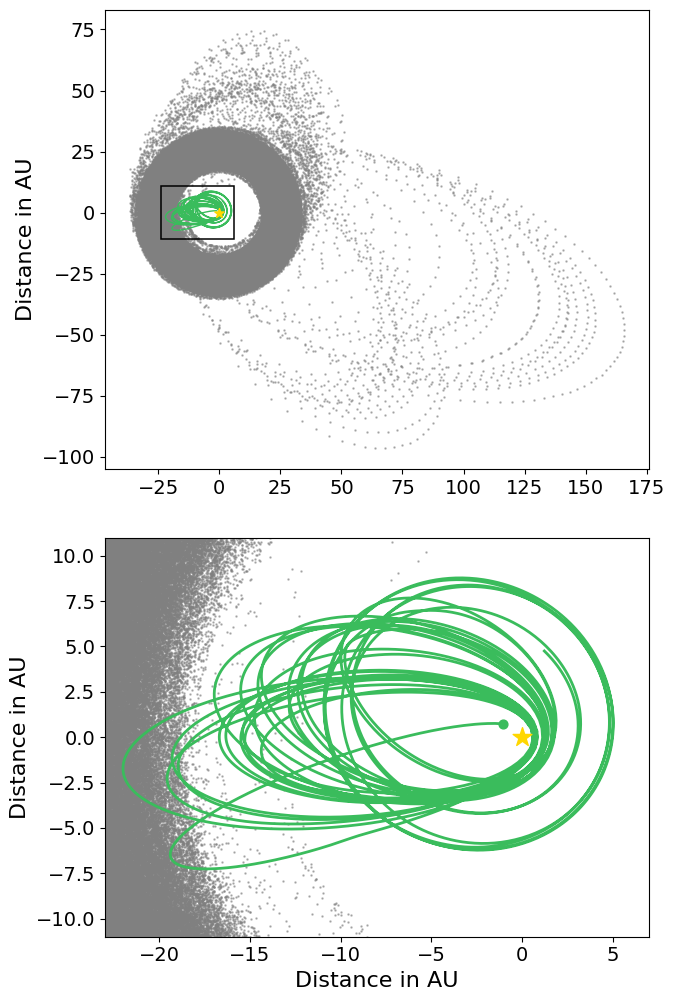}
    \caption{Example of the orbit of a simulated test particle initiated with $a = 25.73$ AU. The gray points show the orbit computed during the low-resolution simulation. The particle initially undergoes interactions with \betapic b, and reaches a periastron distance less than 3 AU at some moment in the simulation. Then the resolution of the simulation is increased, resulting in the orbit shown in green. This high-resolution simulation ends when the periastron distance becomes less than 0.4 AU (marked with the green dot).}
    \label{fig:method_orb}
\end{figure}

\noindent An example showing the resulting orbit of a particle simulated with both the low- and high-resolution time-steps can be found in Fig. \ref{fig:method_orb}. The particle is initiated in the outer system at a semi-major axis of $a = 25.73$ AU and therefore initially simulated with the low-resolution time-step. After reaching the periastron limit, the resolution of this particle is increased in order to resolve the close encounters that are more likely to happen in the vicinity of \betapic c. If a particle was initiated within the 3 AU boundary at the start of the simulation, the particle would be simulated with the high-resolution time-step for its entire lifetime. 

\noindent Test particles were assumed to be mass-less, allowing each particle to be simulated on its own as well as together with other test particles\footnote{In theory, mass-less test particles do not influence one another. However in \rebound, the simulation time-step is adaptive, depending on the proximity of particles to massive particles. This means that removing or adding mass-less particles to a simulation does influence the global outcome, even of the massive particles. This means that two simulations of the same system with different test particles will diverge, despite the fact that test particles do not explicitly exert forces on the massive particles.}. The simulation was divided into smaller instances, containing 25 test particles each, that were then ran in parallel on different CPU cores, greatly increasing the time-efficiency of the simulation. Due to the adaptive step-size used by \verb|IAS15|, particles that get close to a massive body will decrease the step-size, slowing the whole simulation down.  Therefore, to optimize the total integration time further, particles were grouped based on their initial position, with all particles starting within 1 AU being run together, as these are expected to experience similar dynamical evolution pathways.\\

\noindent A significant number of particles may be perpetually captured by either of the planets. Due to the dynamic time-step of the \verb|IAS15| \rebound integrator, captured particles greatly slow the simulation down. We therefore remove particles that reside within the Hill sphere of \betapic c for a continuous time period of more than 0.1 Myr. Out of all test particles in the inner system (< 3 AU), this affects only 0.4\% of all particles.\\

\begin{table}
  \caption[]{Two recent solutions of parameters of the \betapic system: Solution 1 by \citet{Lacour_2021} and solution 2 by \citet{Blunt_2020}, reproduced from \citet{Beust_2024}. We used the parameters from solution 1 to initiate the simulations in \rebound. \citet{Beust_2024} found that these solutions yield equivalent simulation results. Note that our dynamical simulation uses a different reference frame than the sky coordinate system in which these orbital elements are defined, as we define the disk plane to be at an inclination of $0^{\circ}$, so the planets are initialized at inclinations of $90^\circ-i_{b,c}$.}
    \label{tab:initial_conditions}
    \centering
     \begin{tabular}{p{0.15\linewidth}ccc}
        \hline
        \hline
        \noalign{\smallskip}
        Object & Parameter      &Solution 1&  Solution 2\\
        \noalign{\smallskip}
        \hline
        \noalign{\smallskip}
         \boldmath{$\beta$} \textbf{ Pic} &  $ M_*$ $(M_{\odot})$& $1.77$ & $1.85$ \\ \hline
        \boldmath{$\beta$} \textbf{ Pic c} &  $M_c$ $(M_{\text{Jup}})$ & $8.41$& $8.85$ \\
          &  $a_c$ (AU)& $2.70$ & $2.60$ \\
          &  $e_c$ & 0.33 & 0.33\\
         & $i_c$ (deg) & 88.95  & 88.82 \\ 
          &  $\Omega_c$ (deg) & 31.05 & 31.06\\
         &  $\omega_c$ (deg) & 67.70 & 61.02\\
         &  $\tau_c$  & 0.73 & 0.71\\ \hline
         \boldmath{$\beta$} \textbf{ Pic b} &  $M_b$ $(M_{\text{Jup}})$ & $11.73$& $10.00$ \\
         &  $a_b$ (AU)& $9.91$ & $9.95$ \\
          &  $e_b$ & 0.10 & 0.10\\
        & $i_b$ (deg) & 88.99  & 88.98 \\ 
          &  $\Omega_b$ (deg) & 31.81 & 31.81\\
        &  $\omega_b$ (deg) & 196.10 & 202.05\\
         &  $\tau_b$  & 0.71 & 0.73\\
     \end{tabular}

\end{table}

\begin{table}
  \caption[]{Initial conditions of the simulated disk of test particles, modeling planetesimals. All parameters were distributed uniformly. The initial conditions follow simulation set-ups from \citet{RodetLai2024} and \citet{Beust_2024}.}
    \label{tab:initial_conditions2}
    \centering
     \begin{tabular}{p{0.20\linewidth}ccc}
        \hline
        \hline
        \noalign{\smallskip}
        Parameter & Distribution & Unit  \\
        \noalign{\smallskip}
        \hline
        \noalign{\smallskip}
        $a$  & $\mathcal{U}(0.5,45)$ & AU \\ 
        $e$ & $\mathcal{U}(0,0.05)$ & \\ 
        $i$ & $\mathcal{U}(0,2)$ & $^{\circ}$ (deg)\\ 
        $\omega$ & $\mathcal{U}(0,360)$& $^{\circ}$ (deg) \\  
        $l$ & $\mathcal{U}(0,360)$& $^{\circ}$ (deg)
     \end{tabular}

\end{table}

\subsection{Classification}\label{sec:classification}
During the simulation a test particle was assigned a classification if it fulfilled a set of conditions. Firstly, a test particle would be classified as \textit{ejected} if it was found to be on a hyperbolic orbit i.e had an eccentricity $e > 1$ and located at a distance greater than 150 AU, at a distance that is large compared to the orbits of the planets. Secondly, a particle could \textit{collide} with one of the planets in the system. Collisions were evaluated using the \rebound "line" routine that linearly interpolates the positions of a particle between subsequent time-steps \citep{Rein_and_Liu_2012}, following \cite{RodetLai2024}. We tested the sensitivity of the results to the assumed size of the collision radii of the planets, and this is presented in Appendix \ref{sec:Appendix_sensitivity_tests}. Finally, test particles were labeled as \textit{stargrazers} if their periastron distance was below 0.4 AU. This limit is grounded in the work of \cite{Beust_1998} who argue that this is the maximal distance at which dusty particles evaporate sufficiently to produce observable absorption. However, the particle was only classified as a stargrazer if its position was within 1 AU from the star. This additional criterion was necessary to make sure that the particle was no longer interacting with \betapic c, as close encounters with it could momentarily change the periastron distance within the 0.4 AU limit, potentially yielding false classifications. Upon classification in any of these three criteria, the particle was removed from the simulation, and their final orbital elements at that simulation time were saved.

\section{Results}\label{sec:results}
We discuss the outcomes of the simulation in the sections below. Section \ref{sec:branching_ratios} discusses the classification of particles at the end of the simulation, establishing a comparison with \citet{RodetLai2024}. Section \ref{sec:formation_of_exocomets_at_late_times} discusses from which regions particles migrate onto stargrazing orbits in the latter half of the simulation, taken to be representative for the time at the current age of the system. Section \ref{sec:two_families_of_exocomets} evaluates the orbital characteristics of the particles at the moment that they are classified as stargrazers, and derives their radial velocities when seen transiting the line of sight. Here we find that our simulation produced two distinct dynamical families of exocomets. This is in line with the spectroscopic survey of \citet{Kiefer_2014} that demonstrates the existence of two exocomet populations based in part on their separate distributions of observed radial velocities. We therefore propose that the different dynamical families presented by \citet{Kiefer_2014} distinguish particles that originated within the orbits of the planets from those that originated beyond the orbits of the planets. Section \ref{sec:Exocomet_compositions} discusses possible implications for the compositions of the exocomets, as both populations may have been devolitalised to a different degree.

\subsection{Branching ratios}\label{sec:branching_ratios}

\begin{figure*}
    \centering
    \includegraphics[width=1.0\textwidth]{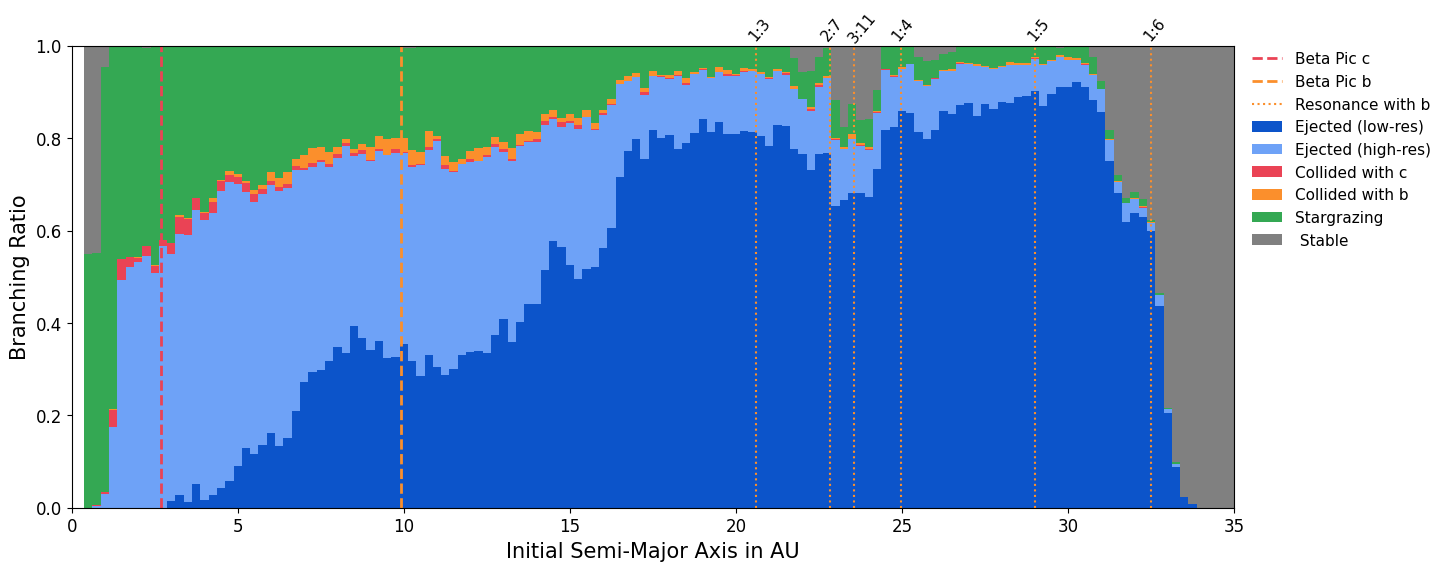}
    \caption{Branching ratios of the entire initial particle population, up to 35 AU after running the simulation (see Methods) for 25 million years. Throughout the system, the majority of particles are ejected (blue) up to a distance of $\sim$ 35 AU, beyond which particles are stable over the system's lifetime. Particles ejected without attaining an orbit that passes within the 3 AU limit and thus only ever simulated using the lower resolution time-step before being ejected are presented in dark blue. However, many particles from the outer system pass within the 3 AU boundary and are also simulated using the high-resolution time-step before being ejected (light blue). All particles initiated on orbital distance smaller than 3 AU are always simulated in high-resolution and therefore also in light blue. A significant number of particles approach the star to within 0.4 AU, which we classify as stargrazers (green). A small number of particles collides with either of the planets. This figure is comparable with the simulation output of \citet{RodetLai2024} (Fig. 19).}
    \label{fig:branchingratio}
\end{figure*}

\begin{figure*}
    \centering
    \includegraphics[width=1.0\textwidth]{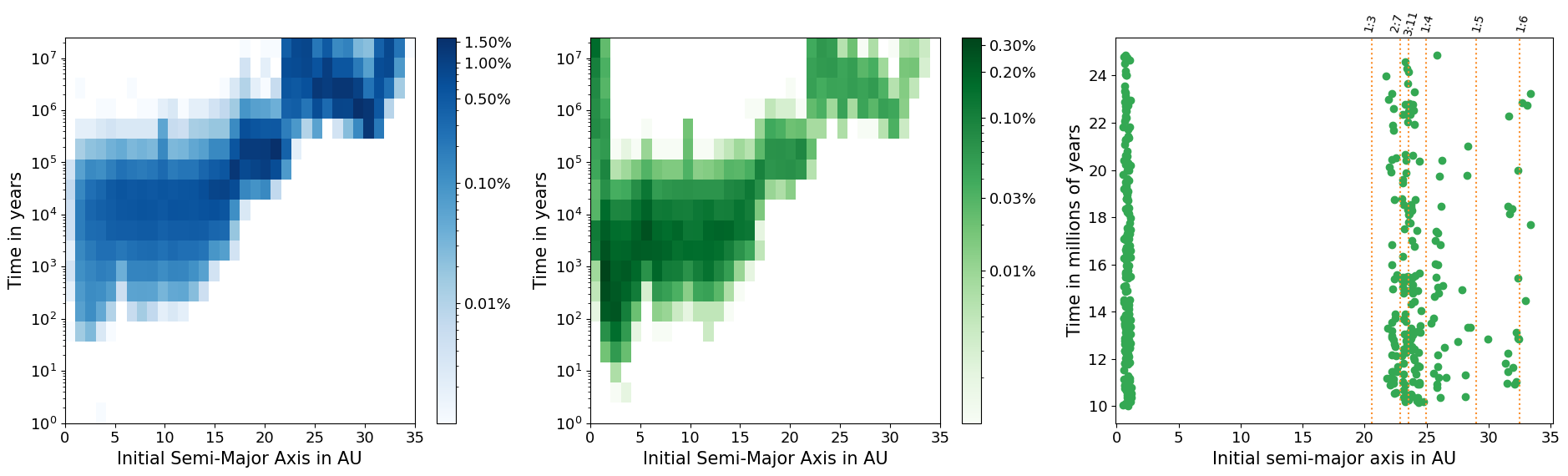}
    \caption{Time-dependency of the evolution of test-particles. The two leftmost panels show the survival times of particles that get ejected (leftmost) or are stargrazing (middle), as a function of the initial semi-major axis at which the particles are seeded. The percentage of all simulated particles that is removed from the simulation is indicated with the colorbar. The rightmost panel shows particles that become classified as stargrazing after 10 million years. We call these "late stargrazers" and investigate them in further detail in section \ref{sec:formation_of_exocomets_at_late_times}. Note that there seem to be two different sources of late stargrazers, in the inner and outer parts of the system respectively. Orbital resonance with \betapic b are also indicated such as in Fig. \ref{fig:branchingratio}, clearly showing that stargrazers are not expected to be produced from low-order resonances (e.g. 1:3 and 1:4), because these have been cleared earlier in the life of the system.}
    \label{fig:lifetimes}
\end{figure*}

\noindent Fig. \ref{fig:branchingratio} shows the relative frequencies (branching ratios) of the classifications of particles at the end of the simulation (see Section \ref{sec:classification}). Fig. \ref{fig:branchingratio} is cut-off at an orbital distance of 35 AU while the simulated disk extends to 45 AU, where all particles remain stable. Our branching ratios are broadly consistent with those presented by \citet{RodetLai2024} (Fig. 19), in that the majority of particles below 35 AU are ejected; while the remainder becomes stargrazing or collides with a planet. We simulated the system with a higher number of particles than \citet{RodetLai2024} to allow for a higher resolving power in the semi-major axis distribution, and this reproduces key properties of the evolution already seen by \citet[][(Fig. 19)]{RodetLai2024} as well as \citet[][Fig. 3]{Beust_2024}: The existence of a region beyond $\sim 35$ AU where the vast majority of particles survive for the duration of the simulation (we refer to this as a "stable" region), as well regions between 20 and 25 AU, where most particles are dynamically cleared within 25 Myr, but a small number on the order of $\sim$ 10\% remains. $15 \%$ of all particles initialized at orbits within 35 AU ultimately evolve to pass within 0.4 AU of the star, and are classified as a stargrazer at some moment in the simulation. The fraction of particles that does so increases from $5.8\%$ for particles initialized between 20 - 25 AU up to $55 \%$ of particles initialized on orbits within 3 AU. The differences between our simulation and that of \citet{RodetLai2024} are that our stargrazer criterion is slightly more inclusive at 0.4 AU versus 0.3 AU of \citet{RodetLai2024}; and that our simulation runs for 25 Myr instead of 10 Myr, which can explain why we find our outer stable region at a larger distance from the star: At later simulation times, the regions between 20 -- 25 AU and beyond 35 AU where significant fractions of particles survived are progressively eroded. As we will show, this erosion is a plausible source for ongoing exocomet creation at the age of the system. Besides stargrazers, the vast majority of particles $75\%$ are ejected from the system (light and dark blue in Fig. \ref{fig:branchingratio}). On the order of 1\% of all planetesimals that are initiated within 35 AU is expected to collide with either of the planets.

\subsection{The formation of exocomets at late times}\label{sec:formation_of_exocomets_at_late_times}
\noindent Although our branching ratios are in line with previous work \citep{RodetLai2024}, we note that branching ratios only show the ultimate outcome of all particles in the simulation, without information about time. Branching ratios therefore do not make evident how stargrazer creation changes during the system's evolution, nor do they show the distribution of stargrazers and ejections being created at the present-day time. Branching ratios also depend on the choice of start and end times of the classification, in our case ranging from 0 to 25 million years\footnote{This means that particles classified as stable, are to be understood to be stable only for the 25 million year duration of the simulation, and not necessarily in the future. Gray areas in the branching ratios (Fig. \ref{fig:branchingratio}) may be referred to as 'stability regions', but this is only valid within the 25 million year time frame.}. Therefore, we also show this classification as a function time in Fig. \ref{fig:lifetimes}. From these figures, it is apparent that the majority of the system within 35 AU is unstable to perturbations by the planets on relatively short ($< 10^5$ yr) timescales, as is well known \citep{RodetLai2024,Beust_2024}. Most of the classified stargrazers thus become stargrazing very early on in the system's lifetime, and these particles are not related to the exocomet phenomenon observed at the present time. However, even after millions of years after the clearing of these inner regions, we find that the creation of stargrazers can continue from three main regions, as seen in the rightmost panel of Fig. \ref{fig:lifetimes}:

\begin{itemize}
    \item the inner edge of the outer stability region close to 35 AU which moves outwards over time,
    \item the intermediate stability regions between 20 to 25 AU that are demarcated by resonances with the orbital period of planet b,
    \item the region within 1.5 AU.
\end{itemize}

\begin{figure}
    \centering
    \includegraphics[width=0.95\linewidth]{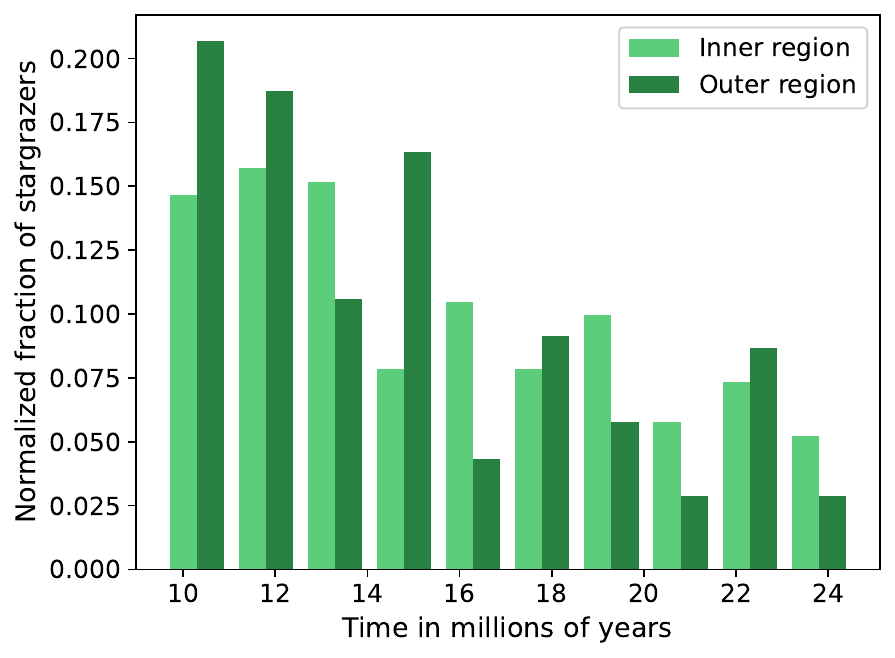}
    \caption{The fraction of particles stargrazing at simulation times greater than 10 million years, normalized by the total number of late stargrazers from each respective stable region. The rate of stargrazers from both regions is decreasing over time at a similar rate.}
    \label{fig:late_stargrazers}
\end{figure}

\noindent To investigate the creation of stargrazers at late times in more detail, we further selected only the particles classified as stargrazers at times greater than 10 Myr from the start of the simulation, which are 399 particles out of a grand total of 133,500. We refer to these as \emph{late stargrazers}. These particles originate from the three main regions listed above, however for the sake of simplicity we group these in two main origins of late stargrazer creation: the \emph{inner region}, which sources the stargrazers produced from the 1.5 AU stability region and the \emph{outer region} which sources late stargrazers from both the intermediate stability regions between 20 and 25 AU, as well as the erosion of the disk beyond $\sim$ 35 AU. As is shown in Fig. \ref{fig:lifetimes} about 52 $\%$, of late stargrazers (208) originate in the outer system and a similar number (191) originates closer to the star. The rates of late stargrazer creation from the inner and outer regions are shown in Fig. \ref{fig:late_stargrazers}. Here, the fraction of particles that stargraze at a given time is normalized to the total number of late stargrazers in each of the two regions. As expected, there is a declining rate of stargrazer creation from both regions over the last 15 million years of the simulation.\\

\noindent Although intuitively, these numbers suggest that the rate of stargrazer formation is roughly equally split between the inner and outer regions, the relative sizes of these stargrazer populations are dependent on the number of particles initialized in each of these regions. As we distributed the particles uniformly in orbital distance (see Methods), the relative rates at which stargrazers are formed cannot be directly inferred from these statistics. We therefore do not derive any conclusions about the absolute number of stargrazer formation, nor the relative rates between the inner and outer populations, besides the observation that the outer region is a plausible source of stargrazers at late times, in addition to the inner region that was already known \citep{Beust_2024}.\\

\subsection{Two families of exocomets?}\label{sec:two_families_of_exocomets}
One of the primary observables from decades of spectroscopic monitoring has been the distribution of radial velocities of the exocomet population, holding clues about their dynamical evolution \citep{Kiefer_2014}. We therefore wish to predict the emergent radial velocity distribution of transiting exocomets generated in our simulation. In particular, we wish to confirm that the inner population is excited via mean-motion resonance with \betapic c, as predicted by \citet{Beust1996} and recently explored in more detail by \citet{Beust_2024}. Both of these objectives are complicated by the fact that the orbit of \betapic c precesses rapidly, with a period of approximately 10,000 years \citep{Beust_2024}. This effectively rotates the directionality of the excitation phenomenon caused by mean-motion resonances \citep[see Fig. 9 in][]{Beust_2024}. In \citet{Beust_2024} this is not considered a problem because they are able to simulate a much larger number of particles because of their use of a symplectic integrator. We only create 399 exocomets and so we do not have the ability to evaluate the statistics of the population at any particular phase of the precession of \betapic c (i.e. the present time). \\

\noindent However, assuming that the generation of exocomets is indeed dominated by interactions with \betapic c, we evaluate the relative orbital alignment of the exocomet populations, compared to the phase of rotation of the semi-major axis of \betapic c. We define the angle $\Delta \omega = \omega - \omega_c$, as is the difference between the argument of periastron of a stargrazing particle and the argument of periastron of \betapic c at the time that the particle is identified as a stargrazer at the moment that it crossed the threshold of 0.4 AU. In defining orbital elements, we use the convention by \citet{Green_1985}, that is also used by \citet{Lacour_2021} who provide the system properties we adopted in this study (see Table. \ref{tab:initial_conditions}). We define a line of sight and a reference plane perpendicular to it. In this coordinate system, one principle axis is along the line of sight, and one principle axis is along the angular momentum vector of the orbit of \betapic c. This means that at the start of the simulation, \betapic c obtains an inclination of 88.95$^\circ$, as observed \citep{Lacour_2021}, but the position angle of the ascending node $\Omega_c$ is equal to zero\footnote{Our reference plane effectively points 31.05 degrees away from North in the sky.}. For each stargrazing particle, we then rotate the euclidean coordinate frame of the simulation by an angle equal to $\Delta \omega$, such that at the epoch of stargrazer creation, the orbit of \betapic c aligns with its orientation on the present day. This allows us to  compare the alignment of stargrazing particles regardless of when they are created in the simulation.\\

\begin{figure*}
    \centering
    \includegraphics[width=\linewidth]{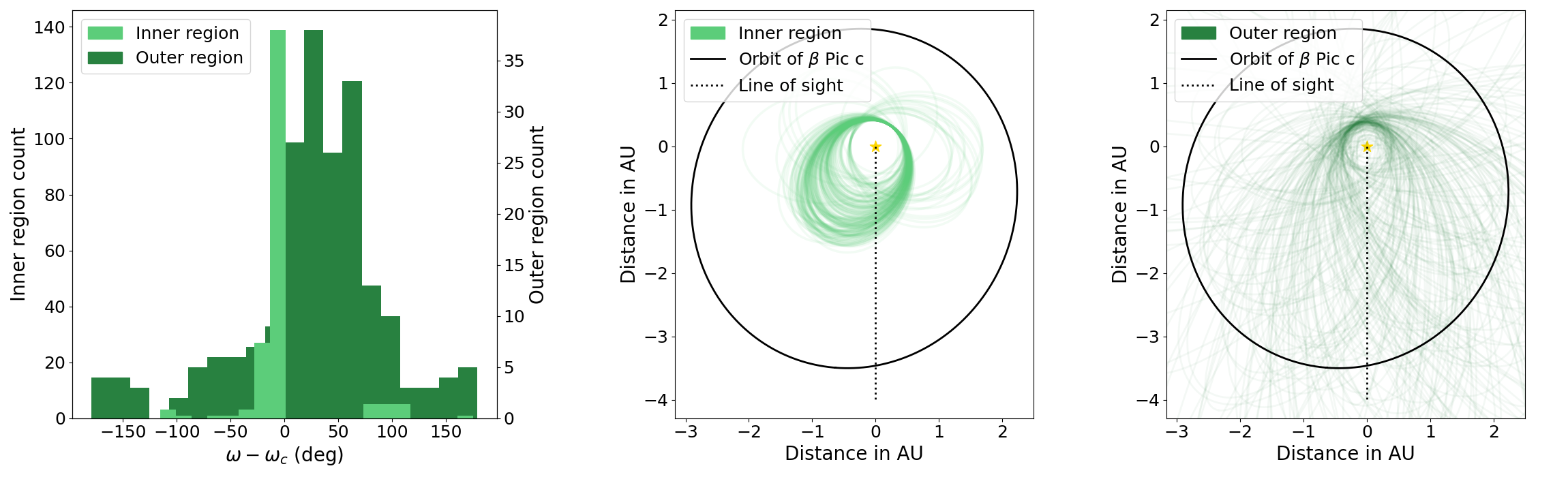}
    \caption{Orbital alignment of the two exocomet populations. The leftmost panel shows the difference between the argument of periastron of the test particle $\omega$ and the argument of periastron of \betapic c $\omega_c$. The two panels to the right show the orbits of stargrazing test particles originating from the inner and outer exocomet reservoirs respectively. The dashed line indicates the direction of the line of sight, and the plane of the sky is parallel to the horizontal axis. Note that in this projection of the coordinate system, particles generally orbit in a counter clock-wise direction.} 
    \label{fig:omegas}
\end{figure*}

\noindent Fig. \ref{fig:omegas} shows the values of $\Delta \omega$ of the exocomets originating from both the inner and the outer regions (see \ref{sec:formation_of_exocomets_at_late_times}). This reveals that the inner population is indeed systematically excited in the direction aligned with the orbital semi-major axis of \betapic c as expected from the mean-motion resonance effect. This therefore confirms the results by \citet{Beust_2024} that mean-motion resonances with \betapic c are capable of generating a distinct dynamical family of exocomets with very similar orbital parameters, hence explaining the observed preferred redshift of the exocomet population. Interestingly, we find that there is also a notable alignment of the orbits of the population of exocomets that originally originate from the outer regions. Intuitively, these may be expected to arrive in the inner system on orbits that are random with respect to the orbit of \betapic c, but interactions with \betapic c do impart a preferential value for the argument of periastron $\omega$.\\

\begin{figure*}
    \centering
    \includegraphics[width=\linewidth]{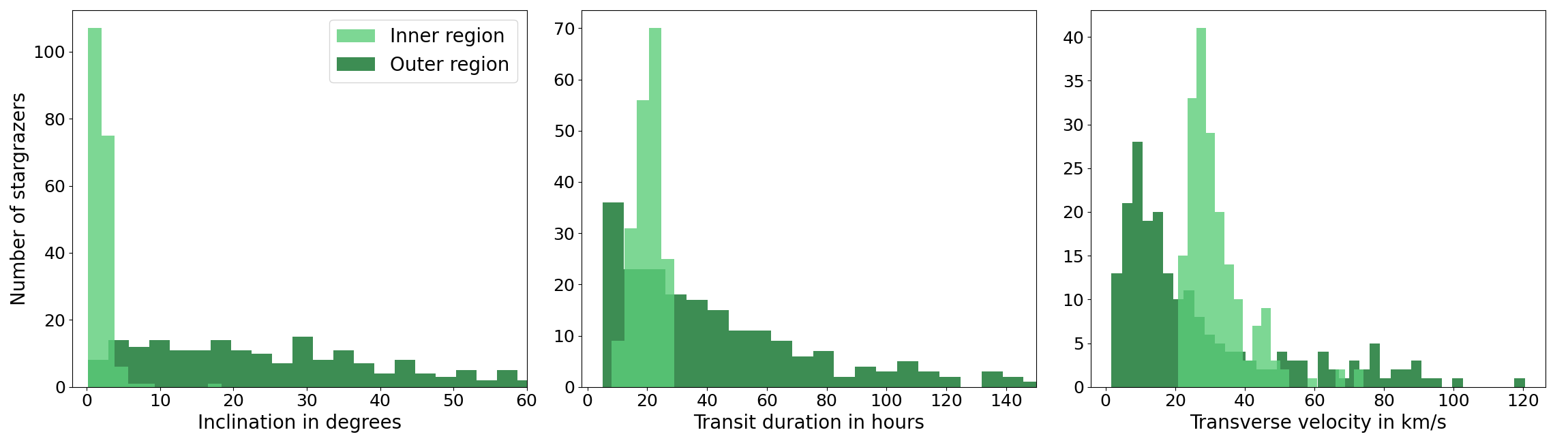}
    \caption{Distributions of exocomet properties for the two identified families. The leftmost panel shows the inclination from edge on orbit ($90^{\circ} - i)$. Note that this panel shows inclinations up to $60^{\circ}$, leaving out approximately $13 \%$ of particles from the outer region with higher inclinations. The middle panel shows the transit duration up to 150 hours (assuming an average impact parameter of 0.5), leaving out approximately 4\% of particles from the outer region with longer transit durations. Finally the rightmost panel shows the transverse velocity.}
    \label{fig:inc+trans}
\end{figure*}

\noindent Having obtained the orbital configuration relative to the present-day orbit of \betapic c, we proceeded to evaluate the radial velocities of the populations during transit. However, stargrazers have non-zero inclinations with respect to the line of sight, so in principle this may only be possible for a small fraction of particles, because many may not transit. The leftmost panel of Fig. \ref{fig:inc+trans} shows the distributions of the inclinations of both the inner and outer populations. The population of particles from the inner region has relatively small\footnote{For simplicity we refer to nearly-edge on orbits as having a small inclination. However, the orbital element \textit{i} is defined to be $90^{\circ}$ for edge-on orbits, so the quantity that is referred to as small is actually $90^{\circ}-i$.} inclinations, and we find that approximately a quarter of them indeed transit. Particles from the outer region have a larger spread of inclinations, and only a small fraction (approximately 2\%) transit. To obtain statistically meaningful samples of radial velocities for both populations, we opt to include particles with impact parameters larger than one. We do this by evaluating the velocity in the direction of the star ($v_{rad}^2 = v_y^2 + v_z^2$), effectively rotating the line of sight out of the plane, to avoid biasing the velocity distribution to smaller values for particles that cross the transit center plane at a large angle. For the particles from the inner region, this creates a negligible bias because the inclinations of the particles are within a few degrees. For the particles in the outer region, including these inclined particles could introduce a bias in the distribution of their radial velocities, but we found that including only low-inclination particles from this population does not significantly change the shape of the distribution (see Fig. \ref{fig:appendix_inc}). We note that the large spread of inclinations of exocomets that originated from the outer system, implies that the edge-on geometry of the \betapic system is not a requirement for observing the exocomet phenomenon, and planetesimal disks with significant inclinations may still produce significant exocomet activity.\\

\begin{figure}
    \centering
    \includegraphics[width=1\linewidth]{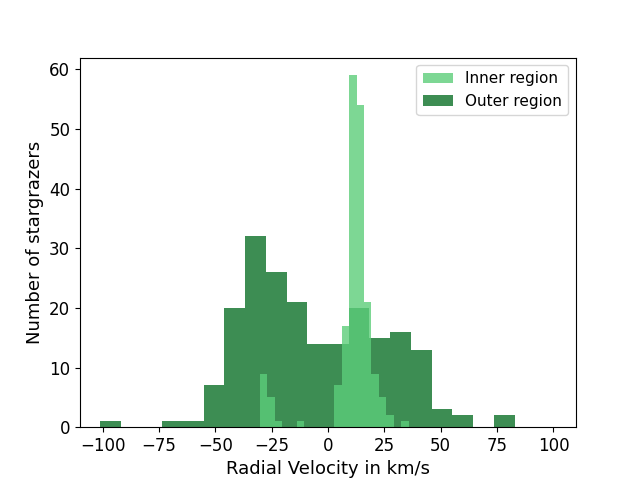}
    \caption{Radial velocity distributions of the two exocomet families. Exocomets originating in the inner region mostly occur in a narrow distribution of radial velocities with a center of 13 km/s and standard deviation of 4 km/s. Exocomets from the outer region have a broader distribution of radial velocities. }
    \label{fig:RV}
\end{figure}

\noindent The resulting distributions of mid-transit radial velocities are shown in Fig. \ref{fig:RV}. The population from the inner region has a narrowly distributed radial velocity that is preferentially redshifted to a mean of 13 km/s with a standard deviation of 4 km/s. The population of particles from the outer regions results in a wider spread of radial velocities. These results are consistent with the observation of \citet{Kiefer_2014} who find exocomets to be distinguished in two dynamical families, based on their radial velocities, transit depths and surface fill fractions. The average radial velocity of inner region particles in our simulation of 13 km/s is consistent with their observed value of $15 \pm 6$ km/s, as well as with the simulation results by \citet{Beust_2024}.\\

\noindent Fig. \ref{fig:inc+trans} also shows the transverse velocities and resulting transit durations (assuming an average impact parameter of 0.5) derived in a similar manner as the radial velocities. We find that transits of both populations typically last for approximately 18 hours, explaining why spectra obtained during a single night tend to appear stable \citep{Kiefer_2014}, apart from a small fraction that are observed to accelerate during transit \citep{Kennedy2018,Hoeijmakers2025}. Transverse velocities of the particles from the inner region peak approximately 30 km/s, and are cut off below 20 km/s. The existence of a lower bound to the transverse velocity is expected because the population of inner particles is confined to a maximum orbital distance that coincides with the inner boundary of the instability region caused by \betapic c. Particles from the outer region have a wider range of transverse velocities, and this is expected because they enter the inner system on a diverse range of orbits. Their distribution peaks at approximately 10 km/s, as these are particles that generally tend to have larger semi-major axes (see Fig. \ref{fig:omegas}). The deep transit event observed by \citet{Zieba2019} allowed for a direct measurement of the transverse velocity of $19.6\pm0.1$ km/s, slightly below the cutoff velocity of the population of particles in our simulation.  This may indicate that this was a particle originally originating from the outer region.\\

\subsection{Simulation limitations}
Although our simulation seems to accurately reproduce the observed radial velocity distribution observed for the population of mean-motion resonance exocomets \citep{Kiefer_2014,Beust_2024}, the wider distribution of the particles originating in the outer system does not exactly match observations, as there appear to be relatively more blue-shifted particles. We point out four important caveats that arise from choices made in our simulation and analysis, that may especially affect the shape of the distribution of this population. \\

\noindent First of all, our simulation results apply to stargrazers on orbits for which the periastron distance has crossed 0.4 AU for the first time. Our simulation therefore does not capture further orbital evolution, that could lead to orbits that get closer the star. This would generally allow for faster exocomets, and this explains why the broad distribution of particles originating from the outer system is limited to radial velocities of $\pm$ 50 km/s, while in reality, much higher radial velocities are also observed \citep{Kiefer_2014}.  Intuitively, we predict that continued orbital evolution of inner-region exocomets will be less pronounced than those from the outer regions, as the latter interact more chaotically with \betapic c. Conversely, it may also be possible that exocomets already become visible when their periastron is further away than 0.4 AU from the star, potentially leading to a lower cut-off value for the transverse velocity of the inner population (Fig. \ref{fig:omegas}).\\

\noindent Secondly, we note that our simulation counts numbers of stargrazing particles. We do not model the possibility that a single exocomet can be observed to transit many times, or that a single particle could break up into a population of particles on similar orbits \citep{Beust1996b}, and this would be required to do a reliable comparison between our velocity distributions and those observed. Doing so would require accurate modeling of tidal break-up, and also the disappearance of exocomets due to sublimation, and such results would be highly model-dependent. However we do expect the distribution of the exocomets from the inner system to be largely robust, because the population of orbits is very coherent.\\

\noindent Thirdly, we note that a roughly equal number of exocomets originated from both regions in our simulation, but that this does not imply that our simulation predicts that these regions should source stargrazers at a relatively equal rate in reality. Our simulation was initialized with an arbitrary density of seed particles that may not match the density distribution of particles in the primordial \betapic planetesimal disk. Furthermore, particles arriving from the outer region to the inner region obtain a larger spread of inclinations than the particles that originate from the inner region (Fig. \ref{fig:inc+trans}), meaning that a relatively smaller number of particles from the outer system become observable as transiting exocomets. Interpreting the observed relative numbers of particles in both populations requires that these effects are robustly taken into account, but this is presently complicated by the first and second caveats already mentioned above.\\

\noindent Fourthly, in correcting for the effect of \betapic c's precessing orbit, we have rotated the coordinate frame for all stargrazing particles to match \betapic c's current orbit. If \betapic b would impart any preferred directionality to the orbits of the exocomet population, this would be washed out. However, we note that \betapic b has a much smaller eccentricity, and this would only affect the population of exocomets coming from the outer regions.\\

\noindent A final note is that our simulation currently uses the IAS15 integrator, but during the preparation of this study a new hybrid symplectic integrator \texttt{TRACE} was released \citep{Lu2024}. \texttt{TRACE} is designed to handle close-encounters by dynamically switching to non-symplectic integration when needed, while maintaining high-speed non-symplectic integration for particles not undergoing close-encounters. In theory, implementing \texttt{TRACE} for the \betapic system may allow for significant increases in simulation speed, and therefore a larger number of simulated particles.

\subsection{Implications for exocomet composition}\label{sec:Exocomet_compositions}

Exocomets are primarily observed via the strong Ca II H\&K lines that are the result of photo-ionization of sublimated refractory dust when the planetesimal is very close to the star. A key question has been whether the transiting exocomets are dominated by volatiles or are mostly rocky \citep{Beust_1989,Beust_1990,Wilson2017, Beust_2024, Kenworthy2025}. Calcium is sublimated from refractory material on the exocomets' surface, but could be part of a volatile-rich outflow that is spectroscopically invisible. Past models of exocomet cloud morphology have assumed icy compositions for the evaporating exocomets resulting in volatile-rich outflows \citep[e.g.][]{Kiefer_2014}. \\

\begin{figure}
    \centering
    \includegraphics[width=\linewidth]{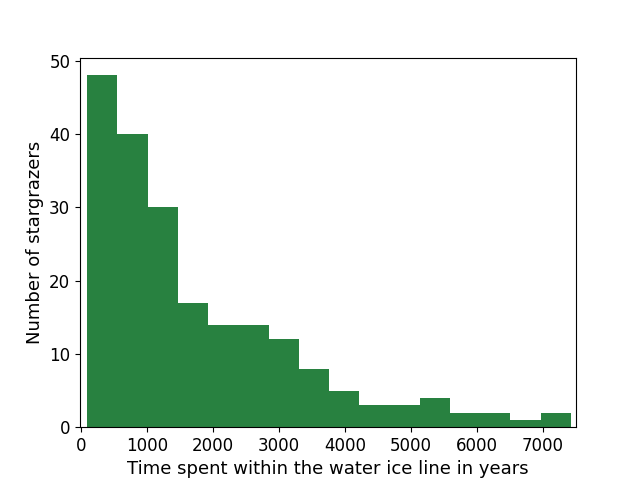}
    \caption{The amount of time that exocomets from the outer region spend inside the ice line (estimated to lie at 8.2 AU) while migrating toward the vicinity of the star before being classified as an exocomet. All of these particles originate outside of the water ice line of the \betapic system (see Fig. \ref{fig:lifetimes}).}
    \label{fig:cooking-time}
\end{figure}

\noindent In our simulation, the exocomet progenitors from the outer region originate at orbital distances much beyond the water ice line located at approximately 8 AU, as determined by matching the sublimation temperature \citep{Lodders2003} to the equilibrium temperature in the \betapic system. We find that exocomet progenitors that get perturbed to become exocomets, typically spend $10^2$ to $10^3$ years within a distance of 8 AU before finally being identified as exocomet candidates at a distance of 0.4 AU, see Fig. \ref{fig:cooking-time}. \citet{Karmann2001} systematically modeled the sublimation of ice-dominated exocomet progenitors, depending on orbital distance, nucleus size and ice content. For ice contained in porous rocky bodies, they find that ice may survive for $>10^5$ years 1 km-sized bodies at orbital distances closer than 5 AU from the star \citep[Fig. 7 in ][]{Karmann2001}. We therefore conclude that exocomets originating from the outer system may have retained significant fractions of ice, that may influence their outgassing characteristics and outgassing composition.\\

\noindent We note that \citet{Kiefer_2014} already observed systematic differences in cloud morphologies between the two populations. The comets with a wide distribution of radial velocities (S family) tends to small fill fractions of the stellar disk. \citet{Kiefer_2014} interpreted that the orbital characteristics of this population are consistent with mean-motion resonance, and that the small fill fraction is caused by a lower evaporation rate, consistent with the expectation that such a population would be devolitalized. However, \citet{Kiefer_2014} predates the discovery of \betapic c and more recent simulations by \citet{Beust_2024} and also our study, that indicate that the other population, with a narrow radial velocity distribution  (D family) is in mean-motion resonance with \betapic c instead. \citet{Kiefer_2014} proposed that the D-family is volatile-rich, but our simulations as well as the simulations of \citet{Beust_2024} demand that this population has been located interior to the orbit of \betapic c for millions of years, meaning that these are expected to be primordially devolitalized -- unless the rocky material would be highly carbonaceous \citep{Karmann2001,Fernandez2006,Beust_2024}. A recent study \citep{Vrignaud2026} using measurements of the population levels of exocomet cloud species, finds that several exocomets associated with family D  transit at significantly larger distances than previously assumed in e.g. \citet{Kiefer_2014}. This is consistent with our simulations that predict that in-transit distances of up to 1.5 AU are expected for the population in mean-motion resonance (see Fig. \ref{fig:omegas}), although one object identified by \citet{Vrignaud2026} transits at a distance of almost 5 AU from the star. Our simulation does not predict any exocomets from the population in mean-motion resonance to reach such large distances, implying that this object is expected to have originated from the regions beyond the orbit of \betapic b. \\

\noindent Our results suggest that the population of comets with a broad radial velocity distribution (S family) originate from the outer system and may hence be richer in volatiles. Further modeling of outgassing processes and cloud dynamics are required to coherently explain the observed cloud properties with these dynamical evolution scenarios. We propose that spectroscopic observations may be capable of probing differences in the volatile contents between these populations. Atomic hydrogen has been observed and proposed to be caused by water ice evaporation of infalling exocomets \citep{Wilson2017}, and an initial attempt to discover CN was recently made by \citet{Kenworthy2025}.\\

\section{Conclusions}
We carried out a sequence of dynamical simulations of over 100,000 particles in the \betapic system using \rebound to explore the dynamical origins of exocomets observed via spectroscopy and photometry \citep{Ferlet1987,Kiefer_2014,Zieba2019}. We assumed a flat disk ($i = \pm 2^{\circ}$) of mass-less test particles initiated on near-circular orbits with semi-major axes between 0.5 to 45 AU and integrated these with the high-accuracy IAS15 integrator for a period of 25 Myr equivalent to the age of the system, and found that a small fraction of particles may be evolved onto stargrazing orbits, originating from two main regions: within and exterior to the orbits of the two planets.\\

\noindent Particles from the outer system gradually become excited by interactions with \betapic b, until they experience close encounters. A fraction of these particles is scattered towards the inner system, interacting chaotically with both planets until some reach highly elliptical, stargrazing orbits. In addition, we find that a significant fraction of 24\% of all particles originating from within 1.5 AU remain stable for the lifetime of the system, and particles from within this inner population exhibit mean-motion resonances with \betapic c, that coherently excite their orbits until they stargraze. This phenomenon has been investigated in detail by \citet{Beust_2024}, who have proposed this to be the explanation for the exocomet phenomenon observed in \betapic since the 1980's \citep{Ferlet1987}. \\ 

\noindent We now conclude that particles from both the inner and the outer regions can dynamically evolve to become exocomets. Evaluating orbital parameters and radial velocities reveals that both populations have distinct dynamical properties: Stargrazers from within the orbit of \betapic c that are excited via MMR obtain a narrow radial velocity distribution centered on $13\pm4$ km/s that is consistent with the simulation results of \citet{Beust_2024}, and that reproduces the observed narrow radial velocity distribution of one observed family of exocomets by \citet{Kiefer_2014} that is centered at $15\pm6$ km/s. We also find that the stargrazers that originate from beyond the orbit of \betapic b enter the inner system on more chaotic orbits, resulting in a broad distribution of radial velocities, and we infer that this can explain the population that is spectroscopically observed to have a broad velocity distribution \citep{Kiefer_2014}. The population of stargrazers from the outer regions may have retained significant volatile reservoirs, which may explain differences in cloud morphology that have been observed spectroscopically \citep{Kiefer_2014}, and distinct spectroscopic signatures due to volatile outgassing may be observable for these exocomets. Interestingly, gaseous exocometary clouds were recently observed at transit distances of up to almost 5 AU \citep{Vrignaud2026}. We also find that exocomets originating from the outer region obtain a large spread of orbital inclinations, meaning that exocomets may be expected to be observable in planetesimal disks that are not viewed edge-on, as is $\beta$ Pictoris.\\ 

\noindent The most important limitations of our simulation are that the criterion by which stargrazers are identified has historically been defined via an orbital distance criterion of 0.4 AU \citep[e.g.][]{Beust_2024}, which may be inaccurate. Moreover, our particles are classified only once, and this ignores the possibility of the same particle transiting multiple times, or streams of particles that result from tidal break-up, that is likely an important effect \citep{Beust1996b}, and finally that our prediction for the number of stargrazers depends strongly on the assumed number of particles initialized in the simulation, making the number of particles generated relative to the unknown size of the real particle reservoir.\\

\begin{acknowledgements}
This work was supported by grants from eSSENCE (grant number eSSENCE@LU 9:3), the Swedish National Research Council (project number 2023-05307), The Crafoord foundation and the Royal Physiographic Society of Lund, through The Fund of the Walter Gyllenberg Foundation. We are grateful to Bibiana Prinoth for important discussions regarding sky-projected orbital elements.
\end{acknowledgements}

  \bibliographystyle{aa} 
  \bibliography{references}

\begin{appendix}

\section{Sensitivity tests} \label{sec:Appendix_sensitivity_tests}

We carried out two tests to determine how our results are influenced by choices of initial conditions. First, we tested how the choice of radii of \betapic b and c affects the rate at which particles are predicted to collide with either of the planets, and thereby the branching ratios and creation of stargrazers at late times. The region between 23 and 24 AU was chosen for this test, as it is a region where all the fates of test particles are possible, see Fig. \ref{fig:branchingratio}. The majority of particles in this region are ejected however a smaller fraction stargraze, collide with either of the planets or remain stable for the full duration of the simulation. Of particular interest is whether the number of late stargrazers from this region is significantly affected by the choice of planet radii.\\

\begin{figure*}[h!]
    \centering
    \includegraphics[width=\linewidth]{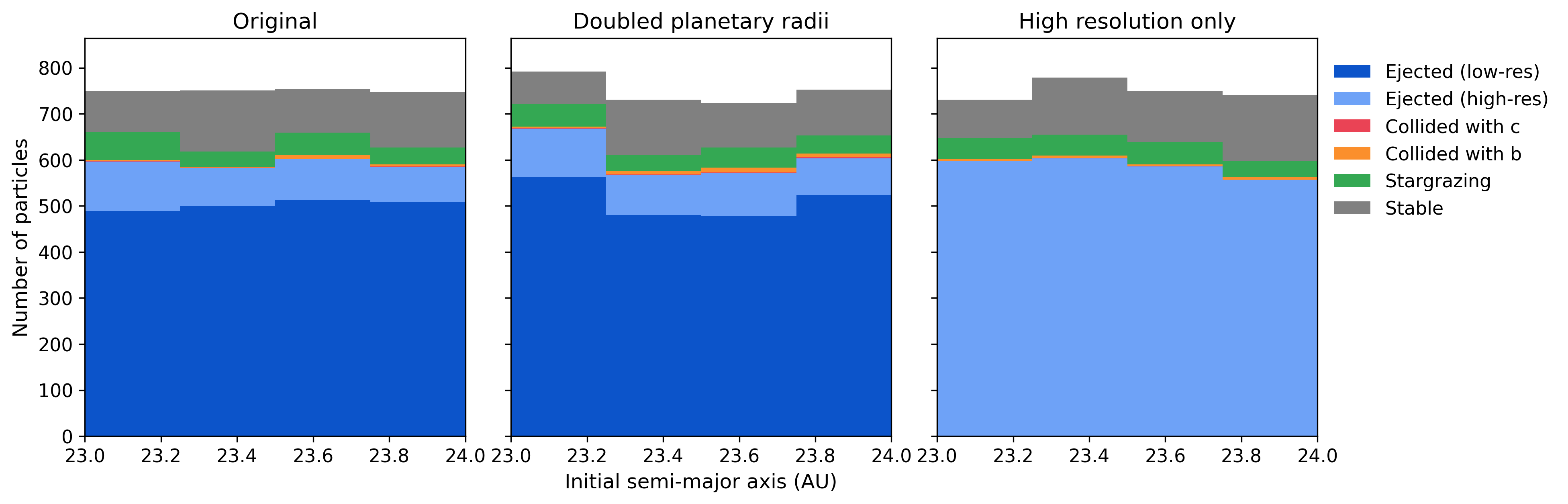}
    \caption{Branching ratios of the region between 23 and 24 AU. The leftmost panel shows a zoom-in of Fig. \ref{fig:branchingratio}, where the simulation was performed using the same set-up as presented in Section \ref{sec:sim_setup}. The middle panel shows the resulting branching ratios of the same region from a simulation where the planetary radii used were doubled. Similarly, the rightmost panel shows the branching ratios of this region resulting from a simulation following the set-up as described in \ref{sec:sim_setup}, but in which all particles were simulated using only the high-resolution timestep.}
    \label{fig:appendix_tests}
\end{figure*}

\noindent We simulated the region between 23 and 24 AU using planetary radii that are twice as large as the ones used in the original simulation, presented in Section \ref{sec:sim_setup}. The resulting branching ratios are shown in the middle panel of Fig. \ref{fig:appendix_tests}. From this sensitivity test we conclude that large changes in the radii of the planets do not significantly alter the results of the simulation: The fraction of colliding particles increases by $70 \%$ and the fraction of particles that remain stable, which decreases by $11 \%$. Importantly, the number of late stargrazers only decreases by one singular late stargrazer, 85 late stargrazers were produced in this region in the original simulation using the nominal radii, and 84 in the test simulation where the radii were doubled. This implies that our uncertainty about the true effective collision radius of the planets, does not meaningfully affect the statistics of the stargazing populations that we present in our conclusions. \\

\noindent Furthermore, we preformed a test of the sensitivity of the choice of maximal time-step using the same range of distances of 23-24 AU. Here, this region was simulated using the high-resolution time-step, that was otherwise only used for particles that cross the 3 AU threshold (see Sec. \ref{sec:sim_setup}). The resulting branching ratios are shown in the rightmost panel of Fig. \ref{fig:appendix_tests}. In this simulation, the total number of stargrazing particles decreases by $3 \%$ and the total number of stable particles increases by $6 \%$ relative to the original simulation. The number of late stargrazers from this region is 75 for the high-resolution test, compared to 85 late stargrazers produced in this region by the original simulation set-up as described in section \ref{sec:sim_setup}. From this we conclude switching the time-resolution does not strongly alter the distribution of outcomes of our simulations, sufficient to violate the conclusion that the particles from regions beyond the orbit of \betapic b constitute a plausible source of stargrazers at the age of the system.

\FloatBarrier

\section{Radial velocity distributions of exocomets with different inclinations} \label{sec:Appendix_RV_inc}
\begin{figure}[h!]
    \centering
    \includegraphics[width=\linewidth]{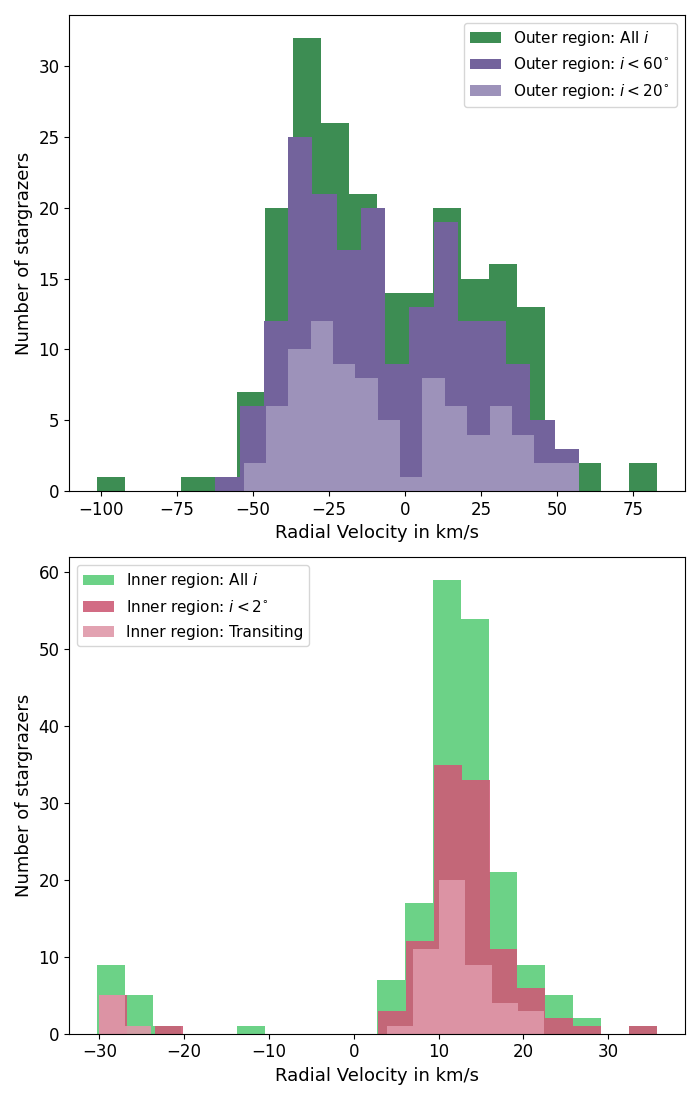}
    \caption{Radial velocity distributions of the inner and outer regions shown for different limits of the inclination.}
    \label{fig:appendix_inc}
\end{figure}

Our radial velocity distributions shown in Fig. \ref{fig:RV} include particles with impact parameters larger than one (see Section \ref{sec:two_families_of_exocomets}). Like discussed in Section \ref{sec:two_families_of_exocomets} this inclusion of inclined stargrazers could introduce a bias to the radial velocity distribution. We test this by limiting the inclinations of the stargrazing particles and compare the radial velocity distributions in Fig. \ref{fig:appendix_inc}. The top panel shows the radial velocity distribution of late stargrazers from the outer population. In green, all late stargrazers are included. In the distributions colored purple, only stargrazers on orbits inclined less than $60^\circ$ (dark purple) and $20^\circ$ (light purple) are included. Similarly, the bottom panel shows the radial velocity distributions of all inner population late stargrazers (green), stargrazers on orbits with inclinations smaller than $2^\circ$ (dark pink) and transiting particles (light pink).\\

\noindent From Fig. \ref{fig:appendix_inc} we conclude that the larger inclinations of late stargrazers do not bias the radial velocity distributions and our conclusions are robust against this.

\end{appendix}

\end{document}